\documentclass[12pt,preprint]{aastex}

\slugcomment{Accepted to ApJ, July 3, 2006}
\newcommand{\etal}{et\,al.}

\begin{document}

\title{Spectropolarimetry of the 3.4 \micron\ Feature in the Diffuse ISM toward
the Galactic Center Quintuplet Cluster}

\author{J.\,E. Chiar\altaffilmark{1}, A.\,J. Adamson\altaffilmark{2}, D.\,C.\,B. Whittet\altaffilmark{3}, 
A. Chrysostomou\altaffilmark{4}, J.\,H. Hough\altaffilmark{4}, \\ T.\,H. Kerr\altaffilmark{2}, 
R.\,E. Mason\altaffilmark{5}, P.\,F. Roche\altaffilmark{6}, and G. Wright\altaffilmark{7}}

\altaffiltext{1}{SETI Institute, 515 N. Whisman Drive, Mountain View, CA
94043, and NASA-Ames Research Center, Mail Stop 245-3, Moffett
Field, CA 94035}

\altaffiltext{2}{Joint Astronomy Centre, 660 N. A'ohoku Place, University Park,
Hilo, Hawaii 96720}

\altaffiltext{3}{Rensselaer Polytechnic Institute, Department of Physics,
Applied Physics and Astronomy, Troy, New York, 12180}

\altaffiltext{4}{Department of Physical Sciences, University of Hertfordshire,
Hatfield, AL10 9AB, UK}

\altaffiltext{5}{Gemini Observatory, 670 N. A'ohoku Place, University Park,
Hilo, Hawaii 96720}

\altaffiltext{6}{Department of Astrophysics, Oxford University, Denys Wilkinson
Building, Keble Road, Oxford OX1 3RH, UK}

\altaffiltext{7}{UK Astronomy Technology Centre, Royal Observatory Edinburgh,
Blackford Hill, Edinburgh EH9 3HJ, UK}

\begin{abstract}
Aliphatic hydrocarbons exhibit an absorption feature at 3.4~\micron\ observed
toward  sources that sample diffuse regions of the interstellar medium. The
absorbers  responsible for this feature are assumed to reside in some component
of interstellar  dust, but the physical nature of the particles (size, shape,
structure, etc.) is  uncertain. Observations of interstellar polarization
provide discrimination.  Since the grains that carry the silicate absorption
feature are known to be  aligned, polarization across the 3.4 \micron\
hydrocarbon feature can be used  to test the silicate core-organic refractory
mantle grain theory.  Although the  3.4 \micron\ feature has been observed to
be devoid of polarization for one line  of sight toward the Galactic center, a
corresponding silicate polarization  measurement for the same line of sight was
not available.  Here, we present  spectropolarimetric observations toward
GCS~3-II and GCS~3-IV toward the Galactic  center, where the 9.7 \micron\
silicate polarization has been previously observed.   We show that polarization
is not detected across the 3.4 \micron\ feature to a  limit of $0.06\pm 0.13$\%
(GCS~3-II) and $0.15\pm 0.31$\% (GCS~3-IV), well below the  lowest available
prediction of polarization on the basis of the core-mantle model.   We conclude
that the hydrocarbons in the diffuse ISM do not reside on the same  grains as
the silicates and likely form a separate population of small grains.
\end{abstract}


\section{Introduction}

Dust grains in the diffuse interstellar medium are composed of both oxygen-rich
and carbon-rich material \nocite{mathis96,draine03} (e.g.\ Mathis 1996; Draine
2003). While the O-rich component is attributed to amorphous silicates, the
composition of the C-rich component remains controversial, proposed candidates
including graphite, diamond, amorphous carbon, aromatic and aliphatic
hydrocarbons, and organic-refractory residues. Of these, aliphatic
hydrocarbons are confirmed to be present in the interstellar media in our own
and other galaxies through observations of vibration absorption features
centered at 3.4 \micron\ (C--H stretch), 6.85 and 7.25 \micron\ (deformation
modes). In the Milky Way, hydrocarbons are detected along lines of sight that
probe the local diffuse ISM within 3~kpc of the Sun
\nocite{adamson_whittet_duley90,sandford_etal91,pendleton_etal94,whittet_etal97}
(Adamson \etal\ 1990; Sandford \etal\ 1991; Pendleton \etal\ 1994; Whittet
\etal\ 1997) and along the line of sight toward the Galactic Center
\nocite{mcfadzean_etal89,sandford_etal91,pendleton_etal94,chiar_etal00,chiar_etal02}
(GC; McFadzean \etal\ 1989; Sandford \etal\ 1991; Pendleton \etal\ 1994; Chiar
\etal\ 2000, 2002).  These same hydrocarbon features are also observed in dusty
Seyfert and ultraluminous infrared galaxies
\nocite{mason_etal04,imanishi_dudley_maloney06,imanishi00agn,spoon_etal04} 
(Imanishi 2000; Mason \etal\ 2004; Spoon \etal\ 2004; Imanishi, Dudley, \&
Maloney 2006).  Within the profile of the 3.4 \micron\ feature, the relative
depths of subfeatures at 3.39, 3.42 and 3.49 \micron\ are indicative of
short-chained hydrocarbons with other perturbing chemical groups (Sandford
\etal\ 1991).  Based on the depth of the corresponding deformation features at
6.85 and 7.25 \micron, the best candidate material is likely to be hydrogenated
amorphous carbon \nocite{chiar_etal00,pendleton_allamandola02,mennella_etal02}
(HAC; Chiar \etal\ 2000; Pendleton \& Allamandola 2002; Mennella \etal\ 2002). 

Although these aliphatic hydrocarbons seem to be ubiquitous constituents of
galactic interstellar media, and much has been learned about their molecular
structure, their formation pathways and physical location within grain
populations are still debated.  Two production routes have been proposed.  One
postulates that organic refractory mantles (ORM) are formed in the diffuse ISM
by UV processing of ice mantles originating in denser clouds
\nocite{greenberg89,greenberg_li96,li_greenberg97} (Greenberg 1989; Greenberg
\& Li 1996; Li \& Greenberg 1997). Dust grains are, indeed, expected to cycle
between dense and diffuse phases of the ISM before being consumed by star
formation \nocite{mckee89} (McKee 1989).  Spectroscopically, carbonaceous
materials produced in this manner provide a reasonable match to the 3.4
\micron\ absorption feature, though they do not reproduce the observed
deformation modes at longer wavelengths (see Pendleton \& Allamandola 2002).
This model predicts that the carrier of the 3.4~\micron\ feature resides in a
mantle on a silicate core.  In the alternative scenario, production occurs in
situ in the diffuse ISM via hydrogen-bombardment of pre-existing carbon
particles \nocite{mennella_etal02} (e.g.\ Mennella et al.\ 2002). This model
predicts hydrocarbons and silicates to be separate grain populations, at least
in low-density environments where grain-grain coagulation is less important.
This difference in physical location predicted for the absorber provides a
means of discrimination.

Much can be learned about the physical properties of dust grains through
polarimetric measurements of their solid-state absorption features.  Excess
polarization across a particular feature is a sign that the grains in which the
absorber resides are aligned by the magnetic field. Furthermore, lack of excess
polarization across a particular feature implies poorly aligned grains or
grains with optical properties, including shape, that induce little or no
polarization. Note that the alignment efficiency is a strong function of
particle size (e.g.\ Draine 2003), with very small ($<0.01~\micron$) grains
being generally much less well aligned compared with larger ``classical''
grains.  The aligned grain population includes silicates, which exhibit excess
polarization across the 9.7 \micron\ and 18.5 \micron\ absorption features in
lines of sight that probe both the diffuse ISM and stars embedded in their
nascent molecular clouds
\nocite{smith_etal00,aitken_etal88_spectropol_gl,aitken_etal89,wright_etal02}
(e.g.\ Aitken \etal\ 1988; Aitken \etal\ 1989; Smith \etal\ 2000; Wright \etal\
2002).  In dense molecular clouds where conditions are sufficiently cold for
ice mantles to form, polarization excesses across the 3 \micron\ H$_2$O-ice
feature \nocite{hough_etal88,kobayashi_etal80,holloway_etal02} (e.g.\ Kobayashi
\etal\ 1980; Hough \etal\ 1988; Holloway \etal\ 2002) and the solid-CO feature
\nocite{chrysostomou_etal96} (Chrysostomou \etal\ 1996) are also detected,
giving firm support for the existence of aligned grains that have silicate
cores surrounded by ice mantles. It is reasonable to suppose that such grains
will continue to produce polarization as they cycle into the diffuse ISM, where
conditions are generally more conducive to efficient alignment
\nocite{lazarian_goodman_myers97} (Lazarian,
Goodman, \& Myers 1997), irrespective of modification to the mantles. 

Since the silicate feature toward the GC is known to be polarized, a good test
of the ORM theory is measurement of polarization across the 3.4 \micron\
feature. If the hydrocarbons reside in mantles on silicate cores, then clearly
the feature should be polarized at a similar level. Nagata \etal\ (1994) found
no significant peak around 3.4 \micron\ in their low-resolution
spectropolarimetric data of several Galactic center sources (including GC-IRS7
and GCS 3-II).  Higher resolution spectropolarimetric observations of  GC-IRS7
were carried out by \nocite{adamson_etal99} Adamson  et al.\ (1999), who showed
that the 3.4 \micron\ feature in GC-IRS7 is not polarized.  This challenged the
ORM theory. However, interpretation of this result was complicated by the fact
that the silicate polarization toward GC-IRS7 was inferred from a measurement
toward the nearby line of sight GC-IRS3.  Although GC-IRS3 is located only
$\sim3\arcsec$  away from GC-IRS7, since the polarization toward GC-IRS7 was
not measured directly, the possibility remains that mantled silicate grains
along this line of sight either fail to align or produce no net polarization
for some other reason (Li \& Greenberg 2002).  To remove this ambiguity, we
have undertaken spectropolarimetry of the 3.4~\micron\ hydrocarbon feature
toward two lines of sight in the Quintuplet cluster for which  silicate
spectropolarimetry already exists \nocite{smith_etal00} (Smith et al.\ 2000). 

In \S2, we describe the observations and the data reduction methods used to
produce the polarization spectra.  In \S3, we discuss the available
spectroscopic and polarimetric observations of the Quintuplet sources and in
\S4 we present the new spectropolarimetric data for the 3.4 \micron\ feature. 
Finally, in \S5 we discuss the implications for grain models of the
non-detection of 3.4 \micron\ polarization.

\section{Observations and Data Reduction}
Using the United Kingdom Infrared Telescope's (UKIRT's) spectrometer CGS4 and
polarimeter, IRPOL2,  we have obtained spectropolarimetry from 3.2 to 3.8
\micron\ of the lines of sight toward GCS 3-II and GCS 3-IV.  The observations
were carried out on the night of 11 August 2003 (UT).  IRPOL2 combines a
Wollaston prism and rotating half-wave plate. Each Stokes Q and U spectrum
resulted from object and sky exposures at each of four position angles of the
half-wave plate.  

Two major issues with these data should be borne in mind. Firstly, the CGS4
spectropolarimetry arrangement is not optimal --- the dispersion due to the
Wollaston prism precedes the slit, and so at least two precise alignments are
necessary to avoid considerable instrumental polarization: (i) alignment
between the slit position angle and Wollaston dispersion direction, (ii)
alignment between the telescope nod and slit position angle. The latter is
straightforward and catered for very precisely by the telescope and instrument
control system. The former requires on-sky measurements in stable conditions to
ensure the slit position angle calibration is adjusted appropriately. In the
case of these observations, conditions were not stable and the target was at a
large negative declination, so the alignment was known to be less than
perfect. In the resulting data there is indeed detectable excess instrumental
polarization compared to that normally expected. There is no evidence that it
contains spectral features, consistent with its origin in throughput losses as
a function of dispersed beam position on the slit. The second issue is the
presence of ripples in the polarization spectrum, in this case having at least
two ``frequencies''. Removal of this ripple was achieved as described by
\nocite{adamson_etal99} Adamson \etal\ (1999), and this appears to have been
successful.

\begin{figure}
\plotone{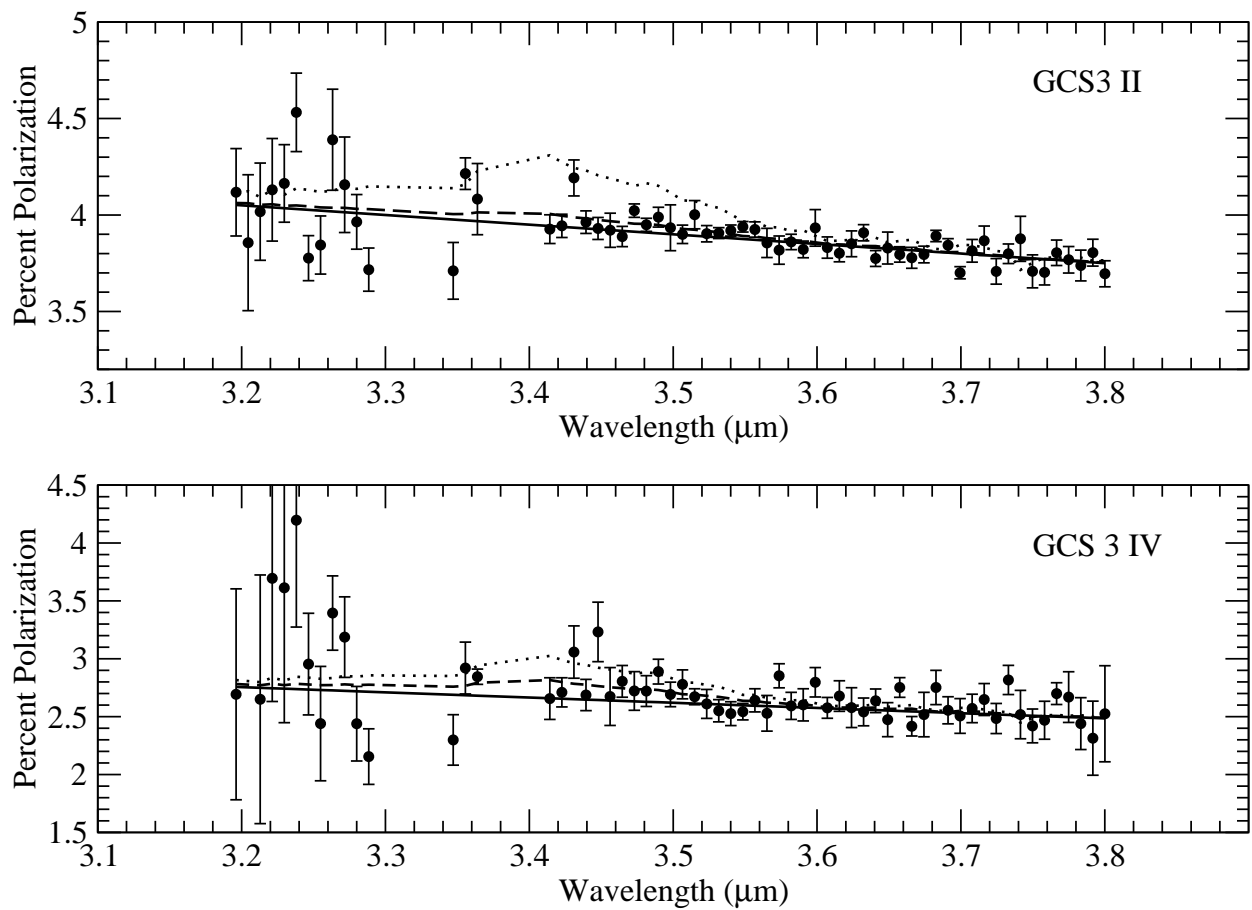}
\caption{Polarization spectra of GCS 3-II [top] and GCS 3-IV [bottom]. 
Observed polarization is shown by the filled circles with error bars.  The
estimated continuum polarization is shown by the solid line. The maximum
allowed polarization as determined from a $\chi^2$ fit, in the 3.4 \micron\
feature is shown by the dashed line. The maximum polarization predicted for
grains with organic mantles, assuming that $\Delta$p/$\tau$ is the same
as for the silicates, is shown by the dotted line.\label{fig:gcs3spec}} 
\end{figure}

Because the polarization in any wavelength bin is a vector quantity, all
binning processes were carried out on the Q and U spectra and the binned
results combined in the usual way to reproduce a binned polarization spectrum. 
The final binned polarization spectra are shown in Fig.~\ref{fig:gcs3spec}. 
The gaps in the observed spectra at 3.29--3.34 and 3.37--3.41 \micron\ occur
where data have been removed because of severe absorption in the
hydrocarbon-containing optical cement in the IRPOL2 Wollaston prism.

\section{The Line of Sight toward the Quintuplet Members GCS 3-II, IV} 

The Quintuplet cluster lies about $14\arcmin$ northeast of the Galactic Center
\nocite{nagata_etal90,okuda_etal90} (Nagata et al.\ 1990; Okuda et al.\ 1990). 
The visual extinction along the line of sight is estimated to be $\sim 29$~mag
\nocite{figer_mclean_morris99} (Figer, McLean, \& Morris 1999).  The line of
sight toward the Quintuplet sources includes both diffuse ISM and dense cloud
material \nocite{schutte_etal98,okuda_etal90,chiar_etal00} (Schutte \etal\
1998; Okuda \etal\ 1990; Chiar et al. 2000). However, the majority of the
absorption arises in the diffuse ISM as evidenced by the observation of weak
3.0 \micron\ H$_2$O-ice features toward GCS 3-I and GCS 3-IV (Chiar \etal\
2000)\footnote{ Based on observations with the Infrared Space Observatory's
(ISO's) Short Wavelength Spectrometer (SWS). The $14\arcsec\times20\arcsec$
ISO-SWS beam  was centered on GCS~3-I and included all four GCS~3 objects.}. 
Assuming a Taurus-like correlation between $A_V$ and $\tau_{3.0}$ (see Whittet
\etal\ 2001), \nocite{whittet_etal01taurusext} we estimate that approximately 6
magnitudes of $A_V$ can be attributed to dense clouds. 

Near-infrared observations of continuum polarization toward the Quintuplet
sources show vectors that lie along the Galactic plane, indicating that it is
interstellar in origin and induced by magnetically aligned dust grains
\nocite{kobayashi_etal83,okuda_etal90} (Kobayashi \etal\ 1983; Okuda \etal\
1990). Silicate feature absorption and corresponding polarization at 9.7
\micron\ is reported toward GCS 3-II and GCS 3-IV by \nocite{smith_etal00}
Smith \etal\ (2000). The $p(\lambda)$ profiles are consistent with pure
absorptive polarization (Smith et al.\ 2000). The position angle of the
silicate polarization observed toward GCS 3-II and GCS 3-IV is essentially
constant across the feature and similar to the interstellar value, and there is
no underlying emission component to complicate the analysis. Therefore, it is
unlikely that the silicate feature has a circumstellar component: it is safe to
assume that it is carried by general-ISM dust along the line of sight. The
9.7~\micron\ features have peak optical depths $\tau=2.9$ and 3.2 in GCS 3-II
and GCS 3-IV, respectively. The corresponding peak polarizations ($p$) are
$9\pm 0.3$ (\%) and $10.2\pm 1.5$ (\%). Based on the observed polarization
spectra, Smith \etal\ (2000) deduce the polarization excess ($\Delta p$) above
the continuum to be 8 (\%) and 8.2 (\%) for GCS 3-II and GCS 3-IV,
respectively.  A measure of the polarization efficiency of the feature is then
given by the ratio $\Delta p/\tau$, which is comparable (2.8 for GCS 3-II and
2.6 for GCS 3-IV) between the two sources. 

The 3.4 \micron\ hydrocarbon absorption feature has been observed toward
several Quintuplet Cluster sources.  Chiar \etal\ (2000) deduce $\tau_{3.4}=0.16$
for GCS 3-I based on spectroscopy with the Short Wavelength Spectrometer of the Infrared Space Observatory.
Chiar, Adamson \& Pendleton 
(in prep.), using higher spatial resolution UKIRT-CSG4 data, find that the
depth of the 3.4 \micron\ feature is spatially invariant.  They measure the
optical depth to be $\tau_{3.4}=0.17$ toward five lines of sight toward the
Quintuplet (including GCS 3-II and GCS 3-IV).   Thus, like the silicates, the
hydrocarbons reside in the diffuse ISM along the line of sight and are unlikely
to be local to the sources themselves.

\section{Spectropolarimetry of the 3.4 \micron\ Hydrocarbon Feature}

Based on the observed optical depth and polarization of the silicate feature, 
it is straightforward to predict that if the 3.4 \micron\ feature is the result
of ORM attached to silicate cores, the 3.4 \micron\ feature should be similarly
polarized with $\Delta p \approx 0.47$\% (GCS 3-II) and 0.44\% (GCS 3-IV). 
Figure 1 shows that the filtered, reconstituted polarization spectra (points
with error bars) of both GCS 3-II and GCS 3-IV are to be essentially
featureless. To estimate the maximum allowable polarization signal in the 3.4
\micron\ feature, a plausible continuum polarization was traced that fits the
observed $p(\lambda)$ spectrum at 3.2 \micron\ and 3.7--3.8 \micron\ (solid
line).  We created a model polarization spectrum using the average hydrocarbon
spectrum calculated by Chiar \etal\ (2002; their Fig.~6) and assuming
$\Delta$p$/\tau$ for the hydrocarbon feature to be equal to that of the
silicate feature.  We then carried out a least squares fit in order to minimize
the residuals between the binned polarization data and model spectrum. The free
parameter was  $\Delta$p, the polarization above continuum at the peak of the
modeled band. The maximum allowable polarization above the continuum is shown
by the dashed line in Fig.~1. The best fits are 0.06$\pm$0.13\% for GCS3-II
and 0.15$\pm$0.31\% for GCS3-IV respectively, consistent with zero in both
cases and well below the predicted level (shown by the dotted line in Fig.~1).
The uncertainties quoted are formal 99\% confidence limits, but ignore possible
systematic errors in continuum placement. The largest uncertainty is probably
in the continuum placement at the short-wavelength end of the spectrum; but
plausible placements of the continuum level result in differences no greater
than 0.02\% in the fitted feature strength and do not therefore substantially
alter the conclusions.

\section{Implications for the Nature of the Hydrocarbon Dust}

We have extended the search for polarization associated with the 3.4 \micron\
absorption feature (Adamson \etal\ 1999) to two additional lines of sight, with
consistent results: the feature is essentially unpolarized. The Adamson \etal\
study made a plausible comparison between their observed limit toward GC IRS7
and the predicted polarization excess for absorbers located on  aligned
silicate grains, but the conclusions were subject to an assumed lack of spatial
variation in the silicate feature, as described in \S1. This ambiguity has now
been removed. Based on models of the polarization properties of  silicate and
carbonaceous analog material, it has been suggested \nocite{li_greenberg02} (Li
\& Greenberg 2002) that polarization across the 3.4~\micron\ feature could be
weaker per unit optical depth compared to the silicate feature.  Li \&
Greenberg (2002) consider mantle/core volume ratios (mantle thicknesses) of
$V_{\rm carb}/V_{\rm sil} = 0.25$, 1, and 2 for spheroids over a range of
elongations.  For the most extreme case of a thick mantle ($V_{\rm carb}/V_{\rm
sil} = 2$), the polarization of the 3.4 \micron\ feature could be up to 25\%
weaker per unit optical depth compared to the silicate feature without
violating the ORM model.  Our data show that the observed polarization across
the 3.4 \micron\ feature is well below even this prediction. 

We conclude that the agent responsible for the hydrocarbon feature in the
diffuse ISM is located in a grain population that is both physically separate
from the silicates and far less efficient as a producer of polarization. A good
candidate is a population of very small carbonaceous grains that are optically
isotropic and/or unresponsive to the alignment mechanism. A likely source of
such grains is indicated by the observations of \nocite{chiar_etal98crl} Chiar
\etal\ (1998), who find a 3.4~\micron\ feature essentially identical to the
interstellar feature in the outflow of a C-rich evolved star. In contrast,
molecular clouds exposed to significant UV flux, presumed to be the formation
sites of ORM, show a conspicuous absence of the 3.4~\micron\ feature
\nocite{shenoy_etal03} (Shenoy \etal\ 2003). Amorphous carbon grains entering
the ISM from the winds of evolved stars will readily become hydrogenated. An
excellent spectroscopic match is obtained to the observed deformation modes, as
well as to the C--H stretch, with laboratory analogs produced by hydrogenation
of nano-sized carbon grains \nocite{chiar_etal00,mennella_etal02} (Chiar \etal\
2000; Mennella \etal\ 2002). Hydrogenated carbon grains are also good
candidates for the 217.5~nm extinction bump, a ubiquitous feature of the
extinction curve in the diffuse ISM \nocite{schnaiter_etal99} (Schnaiter \etal\
1999), which is also observed to be devoid of polarization in all but two (out
of 30) lines of sight studied
\nocite{clayton_etal92,somerville_etal94,clayton_etal95uvpol,wolff_etal97}
(Clayton \etal\ 1992, 1995; Somerville \etal\ 1994; Wolff \etal\ 1997).  

%%%%%%%%%% Something about small grains a la Schnaiter, Furton, etc,   Fix up
%%%%%%%%%% this last paragraph

\acknowledgments
The United Kingdom Infrared Telescope (UKIRT) is operated by the Joint
Astronomy Centre on behalf of the U.K.\ Particle Physics and Astronomy Research
Council.  We thank the Department of Physical Sciences, University of
Hertfordshire for providing IRPOL2 for the UKIRT. J.E.C. was supported by 
NASA's Long Term Space Astrophysics program under grant 399-20-61-02.
D.C.B.W.\ acknowledges financial  support from NASA grant NAG5-13469.

\bibliographystyle{apj}
\bibliography{references}

\begin{thebibliography}{}

\bibitem[\protect\citeauthoryear{{Adamson} et~al.}{{Adamson}
  et~al.}{1999}]{adamson_etal99}
{Adamson}, A.~J., {Whittet}, D. C.~B., {Chrysostomou}, A., {Hough}, J.~H.,
  {Aitken}, D.~K., {Wright}, G.~S.,  \& {Roche}, P.~F. 1999, \apj, 512, 224

\bibitem[\protect\citeauthoryear{{Adamson}, {Whittet}, \& {Duley}}{{Adamson}
  et~al.}{1990}]{adamson_whittet_duley90}
{Adamson}, A.~J., {Whittet}, D. C.~B.,  \& {Duley}, W.~W. 1990, \mnras, 243,
  400

\bibitem[\protect\citeauthoryear{{Aitken} et~al.}{{Aitken}
  et~al.}{1988}]{aitken_etal88_spectropol_gl}
{Aitken}, D.~K., {Smith}, C.~H., {James}, S.~D., {Roche}, P.~F.,  \& {Hough},
  J.~H. 1988, \mnras, 230, 629

\bibitem[\protect\citeauthoryear{{Aitken}, {Smith}, \& {Roche}}{{Aitken}
  et~al.}{1989}]{aitken_etal89}
{Aitken}, D.~K., {Smith}, C.~H.,  \& {Roche}, P.~F. 1989, \mnras, 236, 919

\bibitem[\protect\citeauthoryear{{Chiar} et~al.}{{Chiar}
  et~al.}{2002}]{chiar_etal02}
{Chiar}, J.~E., {Adamson}, A.~J., {Pendleton}, Y.~J., {Whittet}, D.~C.~B.,
  {Caldwell}, D.~A.,  \& {Gibb}, E.~L. 2002, \apj, 570, 198

\bibitem[\protect\citeauthoryear{{Chiar} et~al.}{{Chiar}
  et~al.}{998b}]{chiar_etal98crl}
{Chiar}, J.~E., {Pendleton}, Y.~J., {Geballe}, T.~G.,  \& {Tielens}, A. G.
  G.~M. 1998b, \apj, 507, 281

\bibitem[\protect\citeauthoryear{{Chiar} et~al.}{{Chiar}
  et~al.}{2000}]{chiar_etal00}
{Chiar}, J.~E., {Tielens}, A. G. G.~M., {Whittet}, D. C.~B., {Schutte}, W.~A.,
  {Boogert}, A. C.~A., {Lutz}, D., {van Dishoeck}, E.~F.,  \& {Bernstein},
  M.~P. 2000, \apj, 537, 749

\bibitem[\protect\citeauthoryear{{Chrysostomou} et~al.}{{Chrysostomou}
  et~al.}{1996}]{chrysostomou_etal96}
{Chrysostomou}, A., {Hough}, J.~H., {Whittet}, D. C.~B., {Aitken}, D.~K.,
  {Roche}, P.~F.,  \& {Lazarian}, A. 1996, \apjl, 465, L61

\bibitem[\protect\citeauthoryear{{Clayton} et~al.}{{Clayton}
  et~al.}{1992}]{clayton_etal92}
{Clayton}, G.~C., et~al. 1992, \apjl, 385, L53

\bibitem[\protect\citeauthoryear{{Clayton} et~al.}{{Clayton}
  et~al.}{1995}]{clayton_etal95uvpol}
{Clayton}, G.~C., {Wolff}, M.~J., {Allen}, R.~G.,  \& {Lupie}, O.~L. 1995,
  \apj, 445, 947

\bibitem[\protect\citeauthoryear{{Draine}}{{Draine}}{2003}]{draine03}
{Draine}, B.~T. 2003, \araa, 41, 241

\bibitem[\protect\citeauthoryear{{Figer}, {McLean}, \& {Morris}}{{Figer}
  et~al.}{1999}]{figer_mclean_morris99}
{Figer}, D.~F., {McLean}, I.~S.,  \& {Morris}, M. 1999, \apj, 514, 202

\bibitem[\protect\citeauthoryear{{Greenberg}}{{Greenberg}}{1989}]{greenberg89}
{Greenberg}, J.~M. 1989, in {Interstellar Dust: {IAU} Symposium no. 135}, ed.
  L.~J. {Allamandola} \& A.~G. G.~M. {Tielens} (Dordrecht: Reidel), 345

\bibitem[\protect\citeauthoryear{{Greenberg} \& {Li}}{{Greenberg} \&
  {Li}}{1996}]{greenberg_li96}
{Greenberg}, J.~M.,  \& {Li}, A. 1996, \aap, 309, 258

\bibitem[\protect\citeauthoryear{{Holloway} et~al.}{{Holloway}
  et~al.}{2002}]{holloway_etal02}
{Holloway}, R.~P., {Chrysostomou}, A., {Aitken}, D.~K., {Hough}, J.~H.,  \&
  {McCall}, A. 2002, \mnras, 336, 425

\bibitem[\protect\citeauthoryear{{Hough} et~al.}{{Hough}
  et~al.}{1988}]{hough_etal88}
{Hough}, J.~H., et~al. 1988, \mnras, 230, 107

\bibitem[\protect\citeauthoryear{{Imanishi}}{{Imanishi}}{2000}]{imanishi00agn}
{Imanishi}, M. 2000, \mnras, 319, 331

\bibitem[\protect\citeauthoryear{{Imanishi}, {Dudley}, \& {Maloney}}{{Imanishi}
  et~al.}{2006}]{imanishi_dudley_maloney06}
{Imanishi}, M., {Dudley}, C.~C.,  \& {Maloney}, P.~R. 2006, \apj, 637, 114

\bibitem[\protect\citeauthoryear{{Kobayashi} et~al.}{{Kobayashi}
  et~al.}{1980}]{kobayashi_etal80}
{Kobayashi}, Y., {Kawara}, K., {Sato}, S.,  \& {Okuda}, H. 1980, \pasj, 32, 295

\bibitem[\protect\citeauthoryear{{Kobayashi} et~al.}{{Kobayashi}
  et~al.}{1983}]{kobayashi_etal83}
{Kobayashi}, Y., {Okuda}, H., {Sato}, S., {Jugaku}, J.,  \& {Dyck}, H.~M. 1983,
  \pasj, 35, 101

\bibitem[\protect\citeauthoryear{{Lazarian}, {Goodman}, \& {Myers}}{{Lazarian}
  et~al.}{1997}]{lazarian_goodman_myers97}
{Lazarian}, A., {Goodman}, A.~A.,  \& {Myers}, P.~C. 1997, \apj, 490, 273

\bibitem[\protect\citeauthoryear{{Li} \& {Greenberg}}{{Li} \&
  {Greenberg}}{1997}]{li_greenberg97}
{Li}, A.,  \& {Greenberg}, J.~M. 1997, \aap, 323, 566

\bibitem[\protect\citeauthoryear{{Li} \& {Greenberg}}{{Li} \&
  {Greenberg}}{2002}]{li_greenberg02}
{Li}, A.,  \& {Greenberg}, J.~M. 2002, \apj, 577, 789

\bibitem[\protect\citeauthoryear{{Mason} et~al.}{{Mason}
  et~al.}{2004}]{mason_etal04}
{Mason}, R.~E., {Wright}, G., {Pendleton}, Y.,  \& {Adamson}, A. 2004, \apj,
  613, 770

\bibitem[\protect\citeauthoryear{{Mathis}}{{Mathis}}{1996}]{mathis96}
{Mathis}, J.~S. 1996, \apj, 472, 643

\bibitem[\protect\citeauthoryear{{McFadzean} et~al.}{{McFadzean}
  et~al.}{1989}]{mcfadzean_etal89}
{McFadzean}, A.~D., {Whittet}, D. C.~B., {Bode}, M.~F., {Adamson}, A.~J.,  \&
  {Longmore}, A.~J. 1989, \mnras, 241, 873

\bibitem[\protect\citeauthoryear{{McKee}}{{McKee}}{1989}]{mckee89}
{McKee}, C.~F. 1989, \apj, 345, 782

\bibitem[\protect\citeauthoryear{{Mennella} et~al.}{{Mennella}
  et~al.}{2002}]{mennella_etal02}
{Mennella}, V., {Brucato}, J.~R., {Colangeli}, L.,  \& {Palumbo}, P. 2002,
  \apj, 569, 531

\bibitem[\protect\citeauthoryear{{Nagata} et~al.}{{Nagata}
  et~al.}{1990}]{nagata_etal90}
{Nagata}, T., {Woodward}, C.~E., {Shure}, M., {Pipher}, J.~L.,  \& {Okuda}, H.
  1990, \apj, 351, 83

\bibitem[\protect\citeauthoryear{{Okuda} et~al.}{{Okuda}
  et~al.}{1990}]{okuda_etal90}
{Okuda}, H., et~al. 1990, \apj, 351, 89

\bibitem[\protect\citeauthoryear{{Pendleton} \& {Allamandola}}{{Pendleton} \&
  {Allamandola}}{2002}]{pendleton_allamandola02}
{Pendleton}, Y.~J.,  \& {Allamandola}, L.~J. 2002, \apjs, 138, 75

\bibitem[\protect\citeauthoryear{{Pendleton} et~al.}{{Pendleton}
  et~al.}{1994}]{pendleton_etal94}
{Pendleton}, Y.~J., {Sandford}, S.~A., {Allamandola}, L.~J., {Tielens}, A. G.
  G.~M.,  \& {Sellgren}, K. 1994, \apj, 437, 683

\bibitem[\protect\citeauthoryear{{Sandford} et~al.}{{Sandford}
  et~al.}{1991}]{sandford_etal91}
{Sandford}, S.~A., {Allamandola}, L.~J., {Tielens}, A., {Sellgren}, K.,
  {Tapia}, M.,  \& {Pendleton}, Y. 1991, \apj, 371, 607

\bibitem[\protect\citeauthoryear{{Schnaiter} et~al.}{{Schnaiter}
  et~al.}{1999}]{schnaiter_etal99}
{Schnaiter}, M., {Henning}, T., {Mutschke}, H., {Kohn}, B., {Ehbrecht}, M.,  \&
  {Huisken}, F. 1999, \apj, 519, 687

\bibitem[\protect\citeauthoryear{{Schutte} et~al.}{{Schutte}
  et~al.}{1998}]{schutte_etal98}
{Schutte}, W.~A., et~al. 1998, \aap, 337, 261

\bibitem[\protect\citeauthoryear{{Shenoy} et~al.}{{Shenoy}
  et~al.}{2003}]{shenoy_etal03}
{Shenoy}, S.~S., {Whittet}, D.~C.~B., {Chiar}, J.~E., {Adamson}, A.~J.,
  {Roberge}, W.~G.,  \& {Hassel}, G.~E. 2003, \apj, 591, 962

\bibitem[\protect\citeauthoryear{{Smith} et~al.}{{Smith}
  et~al.}{2000}]{smith_etal00}
{Smith}, C.~H., {Wright}, C.~M., {Aitken}, D.~K., {Roche}, P.~F.,  \& {Hough},
  J.~H. 2000, \mnras, 312, 327

\bibitem[\protect\citeauthoryear{{Somerville} et~al.}{{Somerville}
  et~al.}{1994}]{somerville_etal94}
{Somerville}, W.~B., et~al. 1994, \apjl, 427, L47

\bibitem[\protect\citeauthoryear{{Spoon} et~al.}{{Spoon}
  et~al.}{2004}]{spoon_etal04}
{Spoon}, H.~W.~W., et~al. 2004, \apjs, 154, 184

\bibitem[\protect\citeauthoryear{{Whittet} et~al.}{{Whittet}
  et~al.}{1997}]{whittet_etal97}
{Whittet}, D. C.~B., et~al. 1997, \apj, 490, 729

\bibitem[\protect\citeauthoryear{{Whittet} et~al.}{{Whittet}
  et~al.}{2001}]{whittet_etal01taurusext}
{Whittet}, D.~C.~B., {Gerakines}, P.~A., {Hough}, J.~H.,  \& {Shenoy}, S.~S.
  2001, \apj, 547, 872

\bibitem[\protect\citeauthoryear{{Wolff} et~al.}{{Wolff}
  et~al.}{1997}]{wolff_etal97}
{Wolff}, M.~J., {Clayton}, G.~C., {Kim}, S.-H., {Martin}, P.~G.,  \&
  {Anderson}, C.~M. 1997, \apj, 478, 395

\bibitem[\protect\citeauthoryear{{Wright} et~al.}{{Wright}
  et~al.}{2002}]{wright_etal02}
{Wright}, C.~M., {Aitken}, D.~K., {Smith}, C.~H., {Roche}, P.~F.,  \&
  {Laureijs}, R.~J. 2002, in The Origins of Stars and Planets: The VLT View.
  Proceedings of the ESO Workshop held in Garching, Germany, 24-27 April 2001,
  p. 85., ed. J.~F. {Alves} \& M.~J. {McCaughrean}, 85

\end{thebibliography}

\end{document}